\documentstyle[a4,12pt]{article}
\newcommand{\zbar}{\bar{z}}
\newcommand{\barpartial}{\bar{\partial}}
\newcommand{\be}{\begin{equation}}
\newcommand{\ee}{\end{equation}}
\newcommand{\bea}{\begin{eqnarray}}
\newcommand{\eea}{\end{eqnarray}}
\begin{document}
\begin{titlepage}
\hfill{hep-th/9504096}
\vspace*{\fill}

\centerline{\huge\bf Using Conservation Laws}
\centerline{\ }
\centerline{\huge\bf to Solve}
\centerline{\ }
\centerline{\huge\bf Toda Field Theories}

\vspace{2cm}
\centerline{\large
Erling G. B. Hohler and K\aa re Olaussen}

\vspace{.2cm}
\centerline{\large Institutt for fysikk, NTH,}
\vspace{.1cm}
\centerline{\large Universitetet i Trondheim}
\vspace{.1cm}
\centerline{\large N--7034 Trondheim, Norway.}

\vspace{.5 cm}
\centerline{April 6, 1995}
\vspace*{\fill}
\vspace{.5cm}
\begin{abstract}
We investigate the question of how the knowledge of
sufficiently many local conservation laws for a model
can be utilized to solve the model. We show that for
models where the conservation laws can be written in
one-sided forms, like $\barpartial Q_s = 0$, the problem
can always be reduced to solving a closed system of ordinary
differential equations.
We investigate the $A_1$, $A_2$, and $B_2$ Toda field theories
in considerable detail from this viewpoint.
One of our findings is that there
is in each case a transformation group intrinsic to the model.
This group is built on a specific real form of the Lie algebra used
to label the Toda field theory.
It is the group of
field transformations which leaves the conserved densities invariant.

\end{abstract}

\vspace*{\fill}

\end{titlepage}

\section{Introduction}

For a Hamiltonian system to be ``exactly solvable'' it must have a
sufficient number of conservation laws.
Liouville proved that if a system with $N$ degrees of freedom (i.e.,
with a $2N$-dimensional phase space) has $N$ independent conserved
quantities with mutually vanishing Poisson brackets,
then the system is integrable by quadratures\cite{Arnold}. This means
that there exist (at least in principle) an algorithm for determining
the state of system at time $t$ from the initial data at time
$t_0=0$ (say).

There are (in two space-time dimensions) many non-linear field
theories
which are known to have infinitely many conserved quantities. Since
there
are many ways to count to infinity this fact alone is insufficient to
conclude that the system is integrable; it is however a necessary
condition.
And, even if sufficiently many conservation laws are known, it is in
general quite unclear how they can be used to construct the solution.
The situation is rather the other way around, the existence and form
of the
conservation laws are deduced from an already known method of
constructing the solution, like e.g.\ the inverse scattering method.

In this paper we demonstrate how knowledge of conservation laws can
be
explicitly used to solve e.g.\ the initial value problem, for a
particular
class of field theories.

\subsection{Illustrative example}

The basic idea behind the method is well illustrated by the
(utterly trival) example
\be
  \barpartial\partial\varphi = 0,
  \label{UtterlyTrivial}
\ee
where
$\partial \equiv {\partial}/{\partial t} + {\partial}/{\partial x}
\equiv \partial_z$
and
${\barpartial} \equiv {\partial}/{\partial t}-{\partial}/{\partial x}
\equiv \partial_{\zbar}$.
Eq.\ (\ref{UtterlyTrivial}) has infinitely many conservation laws.
However,
they can all be generated from the two local laws,
\be
  \barpartial J = 0,\qquad \partial{\bar J}=0,
  \label{GeneratingLaw}
\ee
where $J=\partial\varphi=\varphi'+\pi$ and
${\bar J}=\barpartial\varphi=\varphi'-\pi$, as the space integral of
arbitrary polynomials $\cal P$
in $J,\partial J,\partial^2 J,\ldots$, or arbitrary
polynomials $\bar{\cal P}$ in
${\bar J},\barpartial{\bar J},\barpartial^2{\bar J},\ldots$. It
follows from
(\ref{GeneratingLaw}) and the ordinary rules of differentiation that
\be
  \barpartial{\cal P}(J,\partial J,\partial^2 J,\ldots)=0,\qquad
  \partial P({\bar J},\barpartial{\bar J},\barpartial^2{\bar
J},\ldots)=0.
\ee
Thus, all information about the infinite set of conserved quantities
is
already contained in $J=J(x,t)$ and
${\bar J}={\bar J}(x,t)$. This matches the fact that to solve the
problem we must determine
two functions, $\varphi(x,t)$ and $\pi(x,t)\equiv{\dot\varphi}(x,t)$.

At this point we should explain our notation. We do not consider
a canonical formulation in light-cone coordinates, but rather at a
fixed time
$t$. All our conservation laws should thus be viewed as expressions
in terms
of the canonical fields $\varphi$ and $\pi$, with e.g.\
$\partial^n\varphi$ being a short-hand notation for a
more complicated expression.
The latter are found explicitly through repeated use of the
equations of motion, i.e.\ in this example
$\dot\varphi= \pi$, $\dot\pi=\varphi''$.
Here we find for $n\ge1$
\be
  \partial^n\varphi =
  \left(2\partial_x\right)^{n-1}\left(\pi+\varphi'\right),
  \qquad
  \barpartial^n\varphi =
  \left(-2\partial_x\right)^{n-1}\left(\pi-\varphi'\right),
\ee
but with the non-linear equations to be considered later these
substitutions
become more complicated.

Now, to exploit the conservation laws to
solve the initial value problem, we
first determine $J(x,0)=J_0(x)$ and ${\bar J}(x,0)={\bar J}_0(x)$
from
the initial data $\varphi(x,0)$, $\pi(x,0)$. Eq.\
(\ref{GeneratingLaw})
implies a simple time evolution for $J$, $\bar J$:
\be
  J(x,t) = J_0(x+t),\qquad {\bar J}(x,t) = J_0(x-t).
\ee
The fields $\varphi(x,t)$ and $\pi(x,t)$ are now found by inverting
their
relations to the conservation laws, i.e.
\be
    \pi(x,t) = {\frac{1}{2}}\left[J_0(x+t)+{\bar J}_0(x-t)
\right],\quad
    \varphi'(x,t) = {\frac{1}{2}}\left[J_0(x+t)-{\bar J}_0(x-t)
\right].
    \label{inversion}
\ee
Since this is a first order ordinary differential equation for
$\varphi$
the solution is determined only modulo an integration constant
$\varphi_0$.
This is an ambiguity which arises because the conserved currents
$J$ and $\bar J$ are invariant under the field transformation
\[
  \left(\varphi(x,t),\pi(x,t)\right)  \to
  \left(\varphi(x,t)+\varphi_0,\pi(x,t) \right).
\]
Thus, to find the complete solution we need an independent
determination of
say $\varphi(0,t)$.
This can indeed be found by combining (\ref{inversion})
with the equation of motion (\ref{UtterlyTrivial}). Define
$q(t)\equiv\varphi(0,t)$
and $p(t)\equiv\pi(0,t)$. They satisfy the equations of motion
\be
    {\dot q}(t) = p(t),\qquad {\dot p}(t) = \varphi''(0,t).
    \label{t-eqn}
\ee
In general this would not constitute a closed system of equations,
since $\varphi''(0,t)$ is a new unknown quantity.
However, here it may be determined from the conservation laws.
It follows from (\ref{inversion}) that
\be
   \varphi''(0,t) = {\frac{1}{2}}\left[J'_0(t)-{\bar J}'_0(-t)
\right].
\ee
The right hand side of this equation is a known, and may be used to
close the system (\ref{t-eqn}).

The crucial points in the above analysis are (i)~that we have a
sufficient
number of conservation laws which can be written in ``one-sided''
forms
like (\ref{GeneratingLaw}), so that their time development are easily
found, and (ii)~that the relations between the fields and the
conserved
densities are known and can be inverted.
The first condition appears to
be fulfilled for all (conformal) Toda field theories. In this paper
we shall explicitly consider a few of the simplest cases.
The latter relations turn out to form a set of (generally non-linear)
ordinary differential equations, whose order increases
with the number of field components. Thus, the integration of these
equations
introduces a set of undetermined integration constants, reflecting
the
fact that the set of conserved currents are invariant under a certain
group of
transformations on the fields of the model. To find the complete
solution these
integration constants must be determined from initial data. They play
the role
of angle variables in the theory. However, since there is only a
(small) finite
number of such undetermined quantities, most of the (infinite number
of) angle
variables of the model are also determined by the conserved
densities. In this
respect a one-sided local conservation law must contain infinitly
more information than just the (globally) conserved quantities of the
model.

\subsection{Contents and organization of paper}

The rest of this paper is organized as follows:
In the next section we analyse the simplest ($A_1$) Toda field
theory,
i.e.\ the Liouville model, in considerable detail. We
discuss in particular
various methods for solving the initial value problem, but also
show how the conservation of the energy-momentum tensor
can be utilized in a systematic procedure for constructing the
general solution.
In fact,
knowledge of the mechanism behind the general solution makes it
easier to utilize for solving the initial value problem.
We repeat this analysis in the following sections for two Toda field
theories
with two-component fields; in section 3 for the $A_2$ model and in
section 4 for the $B_2$ model. Both of these models have one set of
conservation
laws in addition to conservation of the
energy-momentum tensor. For $A_2$ the additional tensor is of
spin 3, and for $B_2$ it is of spin 4. In both cases we show how the
conservation
laws can be used to derive general solutions of the equations of
motion, and also
indicate practical algorithms for utilizing these general solutions
to solve the
initial value problems.

As indicated above, each Toda field theory is labeled
by the name of a classical Lie algebra. Names were given
originally by Bogoyavlenski\cite{Bogoyavlenski} while giving
the Lax pairs for the models.
This naming scheme is
connected to the method first used by
Leznov and Saveliev\cite{LeznovSaveliev}
to construct general solutions to Toda field theories. Here the
equations
of motion are enforced through a zero curvature condition
for a non-abelian gauge theory constructed from the Lie algebra in
question.
It is not clear from this formulation whether these Lie algebras
have any intrinsic connection to the Toda equations. We show that a
Lie
group obtained by exponentiating one of the real forms of the complex
Lie
algebra is an intrinsic symmetry group for the system. The symmetry
in question
is the group of field transformations which leave the conserved
densitites
invariant. For the $A_1$ Toda field theory this group is $SL(2,R)$,
for
the $A_2$ Toda field theory this group is $SL(3,R)$, and for the
$B_2$ Toda field
theory this group is $Sp(2,R)$. We conclude with a section describing
the
generalization to all Toda field theories.

The Toda field theories have previously been considered by many
authors.
The treatment we feel is most similar to ours is by
Bilal and Gervais\cite{BilalGervais1,BilalGervais2}. Apart from
additional detail, our presentation differs from theirs in
viewpoint and focus. Among other interesting discussions on Toda
field
theories we would like to mention the
references\cite{Mansfield,Palla,Underwood}

\section{The Liouville equation}

The Liouville model can be described by the Lagrangian density
\be
  {\cal L}=\frac{1}{2}(\barpartial\varphi)(\partial\varphi)
           -\sigma\mbox{e}^{\alpha\varphi}.
  \label{LiouLag}
\ee
where $\alpha$ and $\sigma$ are arbitrary real parameters.
One may in the classical case let $\alpha\to1$ and $\sigma\to\pm1$
after appropriate redefinitions of the fields.
Only the case $\sigma=1$ gives a Hamiltonian which is
unbounded from below. The equations of motion become
\be
  \partial\barpartial\varphi=-\sigma \mbox{e}^{\varphi}.
  \label{eq:1.1}
\ee

\subsection{Qualitative behaviour and averaged motion}

We expect the solutions of (\ref{eq:1.1}) to have very
different behaviour when $\sigma=1$ and $\sigma=-1$.
Assume $x$-space to be a circle of unit circumference.
Then (\ref{eq:1.1}) can be thought
to describe a string which is wrapped around a cylinder $S_1\times
R$,
parametrized by the coordinates $(x,\varphi)$,
and can slide along the cylinder. Thus, $\varphi$ measures the
position of the string along the cylinder axis. There is also an
external
potential $\sigma\,\mbox{e}^{\varphi}$ in this direction.
To obtain a qualitative understanding
of the behaviour of the system consider the average position
\[
    {\Phi}(t) = \left\langle \varphi(\cdot,t) \right\rangle
    = \int_0^1 dx\,\varphi(x,t).
\]
With the approximation
\(
    \langle \mbox{e}^{\varphi(\cdot,t)} \rangle
    = \mbox{e}^{\langle \varphi(\cdot,t) \rangle}
\),
which is exact when $\varphi$ is independent of $x$,
we get a closed equation of motion for $\Phi$:
\[
   \frac{d^2}{dt^2}\Phi = -\sigma\mbox{e}^{\Phi}.
\]
This equation has the solutions
\be
   \mbox{e}^{\Phi(t)} =
   \frac{2 a^2}{\cosh^2\left[a(t-t_0)\right]}
   \label{sigmapluss}
\ee
when $\sigma=1$, and
\be
   \mbox{e}^{\Phi(t)} =
   \frac{2 a^2}{\sinh^2\left[a(t-t_0)\right]}
   \label{sigmaminus}
\ee
when $\sigma=-1$.
In the latter case the field $\Phi$ always becomes singular at some
finite time.

When $\sigma=1$
the general (physically interesting)
solution can be written as a fluctuation correction to the average
motion
(\ref{sigmapluss})
\be
   \varphi(x,t) = {\Phi}(t) + \delta\varphi(x,t).
\ee
Since ${\Phi}(t) \to -\infty$ as $\vert t \vert\to\infty$ (which
makes the exponential term in (\ref{eq:1.1}) vanish) it must be
possible
decompose $\delta\varphi(x,t)$ into right- and left-moving waves at
asymptotically early and asymptotically late times,
\bea
  \delta\varphi(x,t) &\sim& \varphi_-(x+t) +
{\bar\varphi}_-(x-t)\quad\mbox{
  as $t\to-\infty$,}\nonumber\\
  \delta\varphi(x,t) &\sim& \varphi_+(x+t) +
{\bar\varphi}_+(x-t)\quad\mbox{
  as $t\to\infty$,}
  \label{decomposition}
\eea
On a cylinder the functions $\varphi_{\pm}$ and ${\bar\varphi}_{\pm}$
must be periodic.

\subsection{The solution of Liouville}

The solution to (\ref{eq:1.1}) was already found by
Liouville\cite{Liouville}.
It can be written in the form
\be
  \mbox{e}^{\varphi(x,t)} = \frac{8\, F'(x+t)\,G'(x-t)}{
   \left[F(x+t) + \sigma\, {\bar G}(x-t) \right]^2}
  \label{LiouSol}
\ee
where $F$ and $G$ are arbitrary functions of their arguments.
Since (\ref{LiouSol}) involves two arbitrary functions it can be
expected to
be the most general solution. However, many choices of $F$ or $G$
lead
to solutions that are unphysical.
Also, it is not entirely trivial to find which
$F$, $G$ correspond to a given set of initial data,
e.g.\ $\varphi(x,0)$ and $\pi(x,0)={\dot\varphi}(x,0)$.
The $x$-independent solutions (\ref{sigmapluss},\ref{sigmaminus})
correspond to the case when $F(\xi)=\exp[a(\xi-t_0)]$ and
$G(\xi)=\exp[a(\xi+t_0)]$.
The asymptotic decompositions (\ref{decomposition})
can easily be achieved by  writing ($\sigma=1$)
\be
  F(x+t) = \mbox{e}^{a(x+t)}\,F_1(x+t),\qquad
  G(x-t) = \mbox{e}^{a(x-t)}\,G_1(x-t).
\ee
By inserting these expressions into (\ref{LiouSol}) we find that
\begin{eqnarray*}
   \mbox{e}^{\varphi} &\sim&
8\mbox{e}^{2at}\,(a\,F_1+F'_1)(a\,G_1+G'_1)/G_1^2\quad
   \mbox{as $t\to-\infty$,}\\
   \mbox{e}^{\varphi} &\sim&
8\mbox{e}^{-2at}\,(a\,F_1+F'_1)(a\,G_1+G'_1)/F_1^2\quad
   \mbox{as $t\to\infty$.}
\end{eqnarray*}
Thus, by choosing $F_1$ and $G_1$ as the periodic solutions to
\be
  F'_1 + a\,F_1 = a\,\mbox{e}^{\varphi_-},
  \qquad
  G'_1 + a\,G_1 = a\,G^2_1\,\mbox{e}^{\bar\varphi_-},
  \label{F_1_and_G_1}
\ee
we find a simple mapping between the fields at asymptotically early
and
asymptotically late times:
\be
  \varphi_+ = \varphi_- -2\, \log F_1, \qquad
  {\bar\varphi}_+ = {\bar\varphi}_- + 2 \log G_1.
  \label{c_transf}
\ee
Since $F_1$ resp.\ $G_1$ are uniquely defined functionals of
$\varphi_-$ resp.\ ${\bar\varphi}_-$, which can be written down
explicitly by solving
(\ref{F_1_and_G_1}),
eqs.\ (\ref{c_transf}) define a canonical transformation which is
the classical analog of the $S$-matrix.

However, for more general given initial data at (say) time $t=0$, it
is not
so clear how $F$ and $G$ should be determined in the most convenient
way.
One (admittedly ugly) solvable system is obtained by differentiating
(\ref{LiouSol}) with respect to $t$ and $x$ before setting $t=0$,
\be
   \frac{F''}{F'} = \frac{2 F'}{(F+\sigma\, G)} +
   {\textstyle\frac{1}{2}}\left(\pi_0+\varphi'_0\right),\qquad
   \frac{G''}{G'} = \frac{2\sigma\,G'}{F+\sigma\, G} -
   {\textstyle\frac{1}{2}}\left(\pi_0-\varphi'_0\right),
   \label{FG-equns}
\ee
with (\ref{LiouSol}) itself as a further (consistent) condition
\be
   F'\,G' = {\textstyle \frac{1}{8}}\,
   \mbox{e}^{\textstyle\varphi_0}\,
   \left(F+\sigma\, G\right)^2.
   \label{cons-cond}
\ee
Note that $F$ and $G$ are not uniquely determined from these
equations; one has
the freedom of choosing 3 arbitrary initial conditions for
(\ref{FG-equns},\ref{cons-cond}). This is related to the fact that
the
representation (\ref{LiouSol}) for $\varphi$ is invariant under a
3-parameter
(M{\"o}bius) group of transformations on $(F,G)$.
This group is generated by the transformations
\be
  \left( F,G \right) \to
  \left(F+a,G+\sigma a \right),\
  \left( F,G \right) \to
  c\,\left(F,G \right),\
  \left(F,G \right) \to \left(F^{-1},G^{-1} \right),
\ee
and may be viewed as a gauge group for the system.

\subsection{Solution by direct use of conservation laws}

Liouville's solution (\ref{LiouSol}) represents a situation where
the mapping from the fields $(\varphi,\pi)$ to the separated
solutions $(F,G)$ is quite complicated, while the inverse mapping is
simple and explicit. As indicated in the introduction, one may
instead
attempt a different route to the solution, utilizing the fact
that the system has conserved currents with simple time development.
In this case the mapping of fields to separated free wave solutions
is simple
and explicit, while the inverse mapping is complicated.
We now investigate this second possibility.

The basic conserved densities for (\ref{eq:1.1}) are the light-cone
components of the conformally improved energy-momentum tensor,
\be
  T \equiv T_{z z} = (\partial\varphi)^2 -2\partial^2\varphi,\qquad
  {\overline T} \equiv T_{\zbar \zbar}=(\barpartial\varphi)^2
  -2\barpartial^2\varphi.
  \label{ImprovedTensor}
\ee
They satisfy $\barpartial T=0$ and $\partial {\bar T}=0$.
And, as with the example of the introduction, any polynomials
${\cal P}(T,\partial T,\partial^2 T,\ldots)$ or
$\overline{\cal P}({\overline T},
\barpartial{\overline T},\barpartial^2{\overline T},\ldots)$
satisfy $\barpartial{\cal P}=\partial\overline{\cal P}=0$. The space
integral
of the corresponding densities will all commute with the Hamiltonian,
and thus be elements in the Lie algebra of the `group of the
Hamiltonian'.
This group has sufficiently many mutually commuting elements to make
the system integrable. However, all the important information is
encoded in
the two fields $T$ and $\overline T$, so these are the quantities one
should
utililize to find the solution.

Expressed by equal time canonical fields we find
\be
  T = \left(\pi+\varphi' \right)^2
  -4\left(\pi'+\varphi'' \right) + 2\sigma\,\mbox{e}^{\varphi},
  \quad
  {\overline T} = \left(\pi-\varphi'\right)^2
  +4\left(\pi'-\varphi''\right) + 2\sigma\,\mbox{e}^{\varphi}.
  \label{EM-tensor}
\ee
There are several ways to read and utilize these equations:
\begin{enumerate}

  \item
  First they can be used to find e.g.\ $T(x,0)=T_0(x)$ and
  ${\overline T}(x,0)={\overline T}_0(x)$ from the initial data,
  $\varphi_0(x)=\varphi(x,0)$ and $\pi_0(x)=\pi(x,0)$.
  From the conservation laws for $T$ and $\overline T$ it
  follows that their time evolution is given as
  $T(x,t)=T_0(x+t)$ and ${\overline T}(x,t)={\overline T}_0(x-t)$.

  \item
  With $T(x,t)$ and ${\overline T}(x,t)$  known and $t$ fixed, eq.\
  (\ref{EM-tensor}) define a set of ordinary differential equations
  in $x$ for $\varphi(x,t)$ and $\pi(x,t)$.
  They are no more complicated than (\ref{FG-equns},\ref{cons-cond}).
  However, the complete solution to these equations requires three
initial
  conditions to be found, say $q(t)\equiv\varphi(0,t)$,
  $r(t)\equiv\varphi'(0,t)$, and $p(t)\equiv\pi(0,t)$.

  \item
  Eqs.\ (\ref{EM-tensor}) may be viewed at fixed $x$ (say $x=0$) as
  a means of expressing $\varphi''(0,t)$ and $\pi'(0,t)$ in terms of
  $(q,r,p)$, $T$, and ${\overline T}$. This may be combined with
  (\ref{eq:1.1}) to find the equations of motion for $(q,r,p)$. They
  become
  \bea
     && {\dot q} = p,\nonumber\\
     && {\dot r} = {\textstyle\frac{1}{2}}pr +
     {\textstyle\frac{1}{8}}\left({\overline T}-T
\right),\label{Initials}\\
     && {\dot p} = {\textstyle\frac{1}{4}}\left(p^2+r^2\right) -
     {\textstyle \frac{1}{2}}\sigma\mbox{e}^{q} -
     {\textstyle\frac{1}{8}}\left({\overline T}+T \right)\nonumber
  \eea
  Here it is understood $T$, $\overline T$ are to be evaluated at
  $x=0$.

\end{enumerate}
In summary, to find the fields at some fixed later time, given the
initial
conditions, one has to solve two sets of ordinary differential
equations in
succession\footnote{Or, more generally, one can solve one set of
equations along
an arbitrary shaped curve in space-time, starting from a point on the
(spacelike)
curve of initial data},
first (\ref{Initials}) and then (\ref{EM-tensor}).
This is still somewhat less efficient than starting from
the known general solution (\ref{LiouSol}).
However, the advantage of this method
is its systematic use of the known conservation laws. Thus, it has
potential for generalization to more complicated equations where the
general solution is unknown.
This method will work as long as a
sufficiently large set of local conservation laws can be explicitly
found.

\subsection{The general solution by ``separation of variables''}

In the previous subsection we indicated how the initial value problem
for the Liouville equation could be solved by systematic use of its
known
conservation laws. The procedure involved the solution of two sets of
ordinary differential equations in succession. In most cases these
equations
must be solved numerically. With present day technology this is in
practice---and as a matter of principle---comparable to the fact that
even
most analytic solutions at some stage must be evaluated numerically
on a computer. (In this respect we still consider the numerical
solution of ordinary
differential equations to be significantly different from the
solution of
{\em partial} differential equations, both because of the
computational
effort involved and the number of parameters in the general
solution.)

However, if we want to compute the fields for many different times,
or derive asymptotic relations like (\ref{c_transf}),
there is still considerable  advantage in working with the general
solution
(\ref{LiouSol}). For this reason we here propose and test
a method for constructing the general solution from the knowledge of
the conservation laws. Our main motivation is of course not to
rederive
(\ref{LiouSol}), but to investigate an ansatz which can be
generalized
to the more complicated Toda field theories.

It is preferable to work with the field $U=\mbox{e}^{-\varphi/2}$.
The energy-momentum tensor (\ref{ImprovedTensor}) becomes
\be
T =
4\,\frac{\partial^2 U}{U},
\qquad
{\overline T}
=
4\,\frac{\barpartial^2 U}{U}.
\label{eq:1.7}
\ee
Thus, the dependency of $U$ on $\zbar$ cancels in $T$,
and its dependency on $z$ cancels in $\overline T$. This
suggests that it should be possible to factor out the
relevant dependencies on $z$ and $\zbar$ explicitly.
To this end we write
\be
U\equiv U(z,\zbar)={\bar u}(z)\, h(z,\zbar)\, u(z) \equiv {\bar
u}\,h\,u,
\label{eq:1.8}
\ee
and make the ansatz that $T$ is a functional of $u$ only, and
$\overline T$ is a functional of $\bar u$ only.
This means that the expressions (\ref{eq:1.7}) for
$T$ resp.\ $\overline T$ should hold with the replacements $U\to u$
resp.\ $U\to {\bar u}$. These are linear $2nd$ order ordinary
differential
equations for $u$ and $\bar u$.
Inserting (\ref{eq:1.8}) into (\ref{eq:1.7}), and using the above
ansatz
imposes two relations on $h$:
\be
   \partial\, u^2 \partial h=0,
   \qquad
   \barpartial\, {\bar u}^2\,\barpartial h = 0.
   \label{eq:1.14}
\ee
It is straightforward to integrate these equations. Define functions
$H=H(z)$ and ${\bar H}={\bar H}(\zbar)$ such that $\partial H =
u^{-2}$
and $\barpartial{\bar H}={\bar u}^{-2}$. Then we obtain the
representations
\be
  h = {\bar\alpha}_0 + {\bar\alpha}_1\,H,
  \qquad
  h = \alpha_0 + \alpha_1\,{\bar H},
  \label{eq:1.22}
\ee
where ${\bar\alpha}_i={\bar\alpha}_i(\zbar)$ and
$\alpha_i=\alpha_i(z)$. These two representations are consistent with
each other
if each $\alpha_i$ is a linear combination of the functions $1$ and
$H$.
Introduce the vectors
\be
  \mbox{\boldmath $H$} = (1,H),
\qquad
  \mbox{\boldmath ${\bar H}$} = (1,{\bar H}).
\ee
Then both the linear equations (\ref{eq:1.14})
are satisfied by the representation (the normalization is for later
convenience)
\be
  h =
  2^{-1/2}\,\left(\mbox{\boldmath $\bar H$}\!\cdot\!\mbox{\boldmath
$X$}\!\cdot\!
        \mbox{\boldmath $H$}^t \right),
  \label{Liou-h}
\ee
where {\boldmath $X$} is a $2\times2$ matrix of real constants. We
remind
the reader that our definitions
of $H$ and $\bar H$ imply
that $u=\left(\partial H\right)^{-1/2}$ and
${\bar u}=\left(\barpartial{\bar H}\right)^{-1/2}$. We have thus
found a
complete solution ansatz for $U = {\bar u}\,h\,u$, which now must
be inserted into (\ref{eq:1.1}). This gives
\[
   h\,\barpartial\partial h - \barpartial h\,\partial h  =
   \sigma\,\barpartial{\bar H}\,\partial H,
\]
which after insertion of the representation (\ref{Liou-h}) gives the
condition
\[
  \left(\mbox{\boldmath $\bar H$}\!\cdot\!
        \mbox{\boldmath $X$}\!\cdot\!
        \mbox{\boldmath $H$}^t
  \right)
  \left(\mbox{\boldmath $\bar h$}\!\cdot\!
        \mbox{\boldmath $X$}\!\cdot\!
        \mbox{\boldmath $h$}^t
  \right)-
  \left(\mbox{\boldmath $\bar H$}\!\cdot\!
        \mbox{\boldmath $X$}\!\cdot\!
        \mbox{\boldmath $h$}^t
  \right)
  \left(\mbox{\boldmath $\bar h$}\!\cdot\!
        \mbox{\boldmath $X$}\!\cdot\!
        \mbox{\boldmath $H$}^t
  \right)
  = \sigma,
\]
with vectors $\mbox{\boldmath $h$}=\mbox{\boldmath $\bar h$}=(0,1)$.
On evaluating the left hand side this reduces to
\be
    \det\mbox{\boldmath $X$}=\sigma.
    \label{LiouX-soln}
\ee
With this condition on {\boldmath $X$}, the conservation laws
together with the
ansatz (\ref{eq:1.8}) have indeed lead us to the general
solution of the Liouville equation:
\be
    \mbox{e}^{\varphi} = \frac{2\,\partial H\,\barpartial{\bar H}}{
    \left(\mbox{\boldmath $\bar H$}\!\cdot\!
          \mbox{\boldmath $X$}\!\cdot\!
          \mbox{\boldmath $H$}^t \right)^2}.
    \label{FinalSolution}
\ee
Liouville's solution (\ref{LiouSol}) correspond to the case that
$\mbox{\boldmath $X$}=
{\footnotesize
\left(\begin{array}{cc}0&1\\-\sigma&0\end{array}\right)}$,
$F(x+t)=H(z)$, and $G(x-t)=-{\bar H}(\zbar)$, with $z=(t+x)/2$ and
$\zbar=(t-x)/2$.

\subsection{The transformation group {\boldmath $SL(2,R)$}}

We have found a solution parametrized
by two arbitrary functions in one variable, and (assume $\sigma=1$) a
matrix $\mbox{\boldmath $X$} \in SL(2,R)$.
What is the role of {\boldmath $X$}?
Note that $H$ resp.\ $\bar H$ is determined from $u$ resp.\ $\bar u$,
hence indirectly via (\ref{eq:1.7}) from $T$ resp.\ $\overline T$.
Equations (\ref{eq:1.7}) are $2nd$ order equations for $u$ or $\bar
u$,
and $3rd$ order equations
for $H$ or $\bar H$. Thus, the general solution
involves 3 arbitrary integration constants
for $H$, and 3 arbitrary integration constants for $\bar H$.
It is, for arbitrary $T$ and
$\overline T$, not possible to find explicit expressions for $u$ and
$\bar u$.
However, assuming that some particular solutions
$u_0$ and ${\bar u}_0$ have been found, it follows
by variation of parameters that the general
solutions to (\ref{eq:1.7}) can be written as
\be
     u = \left({\bar a} + {\bar b}\,\int u_0^{-2}\right) u_0,\qquad
     {\bar u} = \left(a + b\,\int {\bar u}_0^{-2}\right) {\bar u}_0.
\ee
These results applied to the definitions of $H$ and $\bar H$
show that when some particular solutions $H_0$ and ${\bar H}_0$ have
been
found the general
3-parameter classes of solutions can be expressed explicitly as
\be
    H_1 = \frac{a_1\,H_0+a_2}{a_3\,H_0 + a_4},
    \qquad
    {\bar H}_1 = \frac{b_1\,H_0+b_2}{b_3\,H_0 + b_4},
   \label{GenRiccatti}
\ee
where we without loss of generality may choose
\(
   a_1\,a_4-a_2\,a_3 = b_1\,b_4-b_2\,b_3 = 1
\).
By inserting (\ref{GenRiccatti}) into (\ref{FinalSolution})
we find that the substitutions
\[
    H_0 \to H_1, \qquad  {\bar H}_0 \to {\bar H}_1,
\]
are equivalent to keeping {\boldmath $H$ and $\bar H$ fixed,
and instead transforming  $X$} as
\be
   \mbox{\boldmath$ X$} \to
    \left(\begin{array}{cc} b_1 & b_3\\ b_2 & b_4
\end{array}\right)\,
    \mbox{\boldmath$ X$}\,
   \left(\begin{array}{cc} a_1 & a_2\\ a_3 & a_4 \end{array}\right)
   \equiv \mbox{\boldmath$ B$}^t\,\mbox{\boldmath$
X$}\,\mbox{\boldmath$ A$}.
\ee
Thus, there is an $SL(2,R)\times SL(2,R)$ group of transformations on
the fields
such that the energy-momentum tensor is kept invariant. This group is
a product of
an $SL(2,R)$ group of ``gauge transformations''
which leaves the physical fields unchanged,
$
 \mbox{\boldmath $ B$}^t\,\mbox{\boldmath $X$}\,
 \mbox{\boldmath $ A$} = \mbox{\boldmath $X$}
$, and an $SL(2,R)$
group which transforms the physical fields without changing the
energy-momentum
tensor. For a fixed {\boldmath $H$ and $\bar H$ the matrix $X$}
parametrizes the equivalence class of field configurations which
gives rise to
the same conserved densities $T$ and $\overline T$.

The Liouville equation can be formulated as a zero curvature
condition
connected to the Lie algebra $A_1$ (which has $sl(2,R)$ as one of its
real
forms). The above discussion show that a real form of the same Lie
algebra occur in
connection with physical symmetries intrinsic to the Liouville
equation.

\subsection{Revisiting the initial value problem.}

The solution (\ref{FinalSolution}) now leads to a clean way of
solving
the initial value problem, with say the initial data given at $t=0$.

\begin{enumerate}

\item From the initial data we compute the conserved
densities $T$ and $\overline T$. These can be found as functions of
$x$,
which at $t=0$
is directly related to $z=x/2$ and $\zbar=-x/2$.

\item We next determine the functions $u$ and $\bar u$
by solving (\ref{eq:1.7}). This will most likely
have to be done numerically. The equations are of $2nd$
order, thus we need two initial conditions for both $u$ and $\bar u$.
It is convenient to pick one point (say $x=0$)
on the initial data curve, and require that
\be
    u(0) = {\bar u}(0)=1,\qquad\partial u(0) =\barpartial{\bar u}(0)
=0.
    \label{u-init}
\ee

\item With $u$ and $\bar u$ completely known we find $H$ and $\bar H$
by
integration. It is convenient to choose the integration constants
such that
\be
    H(0) = {\bar H}(0) = 0.
   \label{H-init1}
\ee
It now follows from (\ref{u-init}) and the definitions of $H$ and
$\bar H$ that
\be
    \partial H(0) = \barpartial{\bar H}(0) = 1.
    \label{H-init2}
\ee

\item It now remains to determine the matrix {\boldmath $X$}. By
inserting
(\ref{u-init}--\ref{H-init2}) into (\ref{FinalSolution}) we
find\footnote{Note that the left hand sides of eqs.\
(\ref{X-equations})
can all be expressed in terms of the initial data $\varphi_0$ and
$\pi_0$.}
\bea
   \left. \mbox{e}^{-\varphi/2}\right|_0
   &=& 2^{-1/2}\,\mbox{\boldmath $X$}_{11},\nonumber\\
   \left.\partial\left( \mbox{e}^{-\varphi/2}\right)\right|_0
   &=& 2^{-1/2}\,\mbox{\boldmath $X$}_{12},\label{X-equations}\\
   \left.\barpartial\left(\mbox{e}^{-\varphi/2}\right)\right|_0
   &=& 2^{-1/2}\,\mbox{\boldmath $X$}_{21},\nonumber
\eea
which together with (\ref{LiouX-soln}) gives
{\boldmath $X$}. By computing
$\left.\barpartial\partial\mbox{e}^{-\varphi/2}\right|_0 =
2^{-1/2}\,\mbox{\boldmath $X$}_{22}$
we again verify that the equation of motion is consistent with
(\ref{LiouX-soln}).

\end{enumerate}

\noindent
This provides a complete determination of all quantities in
(\ref{FinalSolution}),
which now may be used to compute the fields at arbitrary times. The
connection
between the initial data and general solutions of the Liouville
equation have
also been considered by Papadopoulos and Spence\cite{PapaSpence}

\section{Solution of the {\boldmath $A_2$} Toda field theory}

The $A_2$ Toda field theory can be defined by the
Lagrangian\footnote{For a conventional normalization of the kinetic
terms we
will from here on use the definitions
$\partial \equiv ({\partial}/{\partial t} + {\partial}/{\partial
x})/\sqrt{2}
\equiv \partial_z$
and
${\barpartial} \equiv ({\partial}/{\partial t}-{\partial}/{\partial
x})/\sqrt{2}
\equiv \partial_{\zbar}$.
}
\be
    {\cal L} = \barpartial\phi_1\,\partial\phi_1 +
               \barpartial\phi_2\,\partial\phi_2 +
               \barpartial\phi_2\,\partial\phi_3
   -\frac{1}{2}\sigma_1\mbox{e}^{2(\phi_1-\phi_2)}
   -\frac{1}{2}\sigma_2\mbox{e}^{2(\phi_2-\phi_3)},
   \label{A_2-lag}
\ee
where the $\sigma_i$'s are parameters which may be set to $\pm1$
after
appropriate redefinitions of the fields. Only the case
$\sigma_1=\sigma_2=1$ gives a Hamiltonian which is bounded from
below. We
shall assume this case in our further analysis. The equations of
motion simplify
if we introduce new fields
$\varphi=\frac{2}{3}(\phi_1+\phi_2-2\phi_3)$
and $\chi=\frac{2}{3}(2\phi_1-\phi_2-\phi_3)$,
in which case they become
\be
\partial\barpartial\varphi =
-e^{2 \varphi- \chi},
\qquad
\partial\barpartial\chi =
-e^{-\varphi+2 \chi}.
\label{eq:2.2}
\ee
The third field degree of freedom, $\Phi=\phi_1+\phi_2+\phi_3$,
decouples and satisfies
$\barpartial\partial\Phi=0$.
We note that (\ref{eq:2.2}) is
symmetric under the interchange $\varphi \rightleftharpoons \chi$.
The equations are consistent with the assumption $\varphi=\chi=\phi$,
in
which case they reduce to the Liouville equation.

\subsection{General solution}

Equations (\ref{eq:2.2}) allow two basic sets of conserved densities.
With the notation that $\varphi_n\equiv\partial^n\varphi$,
$\chi_n\equiv\partial^n\chi$,
the set corresponding to the
conformally improved energy momentum tensor is
\be
T = \varphi_1^2-\varphi_1\chi_1
     +\chi_1^2 -\varphi_2 - \chi_2,
\label{eq:2.4}
\ee
and the same expression for $\overline T$ with the replacement
$\barpartial \to \partial$. The other set is
two spin 3 conserved
densities\cite{HohlerOlaussen1,HohlerOlaussen2},
\be
Q
=
 2\varphi_1^2 \chi_1
-2\varphi_1   \chi_1^2
-2\varphi_1   \varphi_2
+2\chi_1  \chi_2
+\varphi_1  \chi_2
-\varphi_2  \chi_1
+\varphi_3
-\chi_3,
\label{eq:2.5}
\ee
and the same expression for $\overline Q$ with the replacement
$\barpartial \to \partial$. Note that $(T,\overline T)$ is even
and $(Q,\overline Q)$ is odd under the $\varphi \rightleftharpoons
\chi$
symmetry. This is seen more explicitly if they are expressed by the
even
$\phi=(\varphi+\chi)/2$ and odd $\psi=\varphi-\chi$ fields. With the
notation
that $\phi_n\equiv\partial^n\phi$ and $\psi_n\equiv\partial^n\psi$,
\be
  T = \phi_1^2 - 2\phi_2 + \frac{3}{4}\psi_1^2,\quad
  Q = (2\phi_1^2-\phi_2)\psi_1
  -\frac{1}{2}\psi_1^3 - 3\phi_1\psi_2 + \psi_3.
\ee
These conserved densities satisfy
$\barpartial T = \partial {\overline T}= \barpartial Q =
\partial{\overline Q} = 0$.
Thus, their time evolution is explicitly known.
It is convenient to express them by the
new variables $U=\mbox{e}^{-\varphi}$ and $V=\mbox{e}^{-\chi}$.
With the notation that $U_n\equiv\partial^n U$ and
$V_n\equiv\partial^n V$ they
become
\bea
T&=&
\frac{U_2}{U} + \frac{V_2}{V} -
\frac{U_1 V_1}{U V}
\label{eq:2.12}
\\
Q &=&
\frac{V_3}{V} -\frac{U_3}{U} +
\frac{U_1 V_2 - U_2 V_1}{U\,V} +
\frac{U_2 U_1}{U^2} -
\frac{V_2 V_1}{V^2} +
\frac{U_1 V_1^2}{U\,V^2} -
\frac{U_1^2 V_1}{U^2\,V}.
\label{eq:2.13}
\eea
and the same equations for $\overline T$ and
$\overline Q$ with the replacement $\partial\to\barpartial$. In
solving the initial
value problem these relations should be viewed as equations for the
fields
$U$ and $V$, with $(T,\overline T)$ and $(Q,\overline Q)$ already
known quantities.
We want to write them in separated form. To this end we
use the conserved density (\ref{eq:2.12}) to eliminate
$V_2$ resp.\ $U_2$ in (\ref{eq:2.13}).
We find
\be
  \frac{U_3}{U} - T\,\frac{U_1}{U} =
  \frac{1}{2}(\partial T - Q),\qquad
  \frac{V_3}{V} - T\,\frac{V_1}{V} =
  \frac{1}{2}(\partial T + Q),
  \label{V12-lign}
\ee
which are seen to be linear equations for $U$ and $V$. There is a
similar
set of equations for $U$ amd $V$, obtained by making the
replacement $T\to{\overline T}$,
$Q\to{\overline Q}$, and $\partial\to\barpartial$ in
(\ref{V12-lign}).

As for the Liouville model we try to factor out the relevant
dependencies
on $z$ and $\zbar$ explicitly. We write
\be
   U = {\bar u}(\zbar)\,h(z,\zbar)\,u(z),
   \qquad
   V = {\bar v}(\zbar)\,k(z,\zbar)\,v(z),
\ee
and make the ansatz that
$T,Q$ are functionals of $u$ and $v$
only, and that ${\overline T},{\overline Q}$ are functionals of $\bar
u$ and
$\bar v$ only.
Thus $u,v$ and $\bar u,\bar v$ are determined by ordinary
differential equations;
those for $u,v$ by making the replacement $(U,V)\to (u,v)$ in
(\ref{V12-lign}),
and the similar ones for $\bar u,\bar v$ by making the replacement
$(U,V,T,Q,\partial)\to({\bar u},{\bar v},\overline T,\overline
Q,\barpartial)$
in (\ref{V12-lign}).
We still have to determine $h$ and $k$ from equations in both $z$ and
$\zbar$,
but these will be explicitly given, and turn out to be quite
managable.

By inserting the factorized expressions for the $U$ and $V$ into
(\ref{V12-lign}), and
demanding that $T$ and $Q$ should be the above mentioned
functionals of $u$ and $v$ only, we obtain equations for $h$ and $k$.
They may be cast into the form
\be
    \partial\, \frac{v^2}{u}\,\partial\,\frac{u^2}{v}\,\partial h =0,
    \qquad
    \partial\, \frac{u^2}{v}\,\partial\,\frac{v^2}{u}\,\partial k =0.
    \label{h12-lign}
\ee
There is another set of equations for $h$ and $k$,
obtained by making the replacement
$(u,v,\partial)\to({\bar u},{\bar v},\barpartial)$ in those above.
It is straightforward to
integrate (\ref{h12-lign}).
Define functions $H=H(z)$ and $K=K(z)$ such
that $\partial H = v/u^2$ and $\partial K = u/v^2$. Then
\be
  h = {\bar\alpha}_0 + {\bar\alpha}_1\, H + {\bar\alpha}_2 \int
K\,\partial H,
  \qquad
  k = {\bar\beta}_0 + {\bar\beta}_1\, K + {\bar\beta}_2 \int
H\,\partial K,
  \label{h12-z}
\ee
where all the coefficients may depend on $\zbar$:
$\bar\alpha_0=\bar\alpha_0(\zbar)$ etc.
Likewise, to solve the $\zbar$-dependence
of $h$ and $k$  we define functions ${\bar H}={\bar H}(\zbar)$
and ${\bar K}={\bar K}(\zbar)$ such
that $\barpartial{\bar H}= {\bar v}/{\bar u}^2$ and
$\barpartial {\bar K}={\bar u}/{\bar v}^2$. Then
\be
  h = {\alpha}_0 + {\alpha}_1\, {\bar H} + {\alpha}_2 \int {\bar K}\,
      \barpartial {\bar H},
  \qquad
  k = {\beta}_0 + {\beta}_1\, {\bar K} + {\beta}_2 \int {\bar H}\,
      \barpartial {\bar K},
  \label{h12-zbar}
\ee
where all the coefficients may depend on $z$: $\alpha_0=\alpha_0(z)$
etc.

Equations (\ref{h12-z},\ref{h12-zbar}) are consistent with each other
if each
of the coefficients $\alpha_i$ is a linear combination
of the functions $1$, $H$, $\int K\, \partial H$, and each of the
coefficients $\beta_i$ is a linear combination of the
functions $1$, $K$, $\int H\,\partial K$. Define
\bea
   &\mbox{\boldmath $H$}      = (1,H,\int K\,\partial H),\qquad
   &\mbox{\boldmath $K$}      = (1,K,\int H\,\partial
K),\label{FG-vectors}\\
   &\mbox{\boldmath $\bar H$} = (1,{\bar H},\int {\bar K}\,
    \barpartial {\bar H}),\qquad
   &\mbox{\boldmath $\bar K$} = (1,{\bar K},\int {\bar H}\,
    \barpartial {\bar K}).
\eea
Then we have found that the $h$ and $k$  must be of the form
\be
    h =
    \mbox{\boldmath $\bar{H}$}\!\cdot\!\mbox{\boldmath $X$}\!\cdot\!
    \mbox{\boldmath $H$}^t,
    \qquad
    k = \mbox{\boldmath $\bar{K}$}\!\cdot\!\mbox{\boldmath
$Y$}\!\cdot\!
    \mbox{\boldmath $K$}^t,
    \label{h12-form}
\ee
where {\boldmath $X$ and $Y$} are $3\times 3$ matrices of real
constants.

Now inserting the ansatz $\varphi=-\log(\bar{u}\,h\,u)$ and
$\chi=-\log({\bar v}\,k\,v)$ into the Toda equations (\ref{eq:2.2})
gives
the equations
\(
    h\,\barpartial\partial h - \barpartial h\,\partial h =
    k\, \barpartial {\bar H}\,\partial H
\)
and
\(
    k\,\barpartial\partial k - \barpartial k \partial k =
    h\, \barpartial {\bar K}\,\partial K
\).
To evaluate these expressions further we first compute
\be
  \partial\mbox{\boldmath $H$} =
   \left(0,1,K\right)\,\partial H
   \equiv \mbox{\boldmath $h$}\,\partial H, \quad
  \partial \mbox{\boldmath $K$} =
   \left(0,1,H\right)\,\partial K
   \equiv \mbox{\boldmath $k$}\,\partial K,
\ee
and the similar barred relations.
It then follows from (\ref{h12-form}) that
\begin{eqnarray*}
    &&h \barpartial\partial h=
    \left(\mbox{\boldmath $\bar {H}$}\!\cdot\!\mbox{\boldmath
$X$}\!\cdot\!
    \mbox{\boldmath $H$}^t\right)\,
    \left(\mbox{\boldmath $\bar{h}$}\!\cdot\!\mbox{\boldmath
$X$}\!\cdot\!
    \mbox{\boldmath $h$}^t\right)\,\left(\barpartial{\bar
H}\,\partial H\right),\\&&
    \barpartial h\,\partial h =
    \left(\mbox{\boldmath $\bar h$} \!\cdot\!\mbox{\boldmath
$X$}\!\cdot\!
    \mbox{\boldmath $H$}^t\right)\,
    \left(\mbox{\boldmath $\bar H$}\!\cdot\!\mbox{\boldmath
$X$}\!\cdot\!
    \mbox{\boldmath $h$}^t\right)\,\left(\barpartial{\bar H}\partial
H\right),
\end{eqnarray*}
and the same with
$(h,H,{\bar H},
\mbox{\boldmath  $H$},
\mbox{\boldmath $h$},
\mbox{\boldmath $\bar H$},
\mbox{\boldmath $\bar h$},
\mbox{\boldmath $X$})
\to
(k,K,{\bar K},
\mbox{\boldmath  $K$},
\mbox{\boldmath $k$},
\mbox{\boldmath $\bar K$},
\mbox{\boldmath $\bar k$},
\mbox{\boldmath $Y$})
$.
We obtain the algebraic equations\footnote{If one wants to solve the
model
(\ref{A_2-lag}) with $\sigma_1,\sigma_2\in \left\{1,-1\right\}$ one
should
at this point multiply the right hand side of (\ref{Yij-lign}) by
$\sigma_2$
and the right hand side of (\ref{Xij-lign}) by $\sigma_1$. These
modifications
can be undone by the substitutions
$\mbox{\boldmath $Y$}\to\sigma_2\mbox{\boldmath $Y$}$ and
$\mbox{\boldmath $X$} \to \sigma_1\mbox{\boldmath $X$}$.
}
\bea
   &&\left(\mbox{\boldmath $\bar H$}\!\cdot\!\mbox{\boldmath
$X$}\!\cdot\!
   \mbox{\boldmath $H$}^t\right)\,
   \left(\mbox{\boldmath $\bar{ h}$}\!\cdot\!\mbox{\boldmath
$X$}\!\cdot\!
   \mbox{\boldmath $h$}^t\right) -
   \left(\mbox{\boldmath $\bar{h}$}\!\cdot\!\mbox{\boldmath
$X$}\!\cdot\!
   \mbox{\boldmath $H$}^t\right)\,
   \left(\mbox{\boldmath $\bar{H}$}\!\cdot\!\mbox{\boldmath
$X$}\!\cdot\!
   \mbox{\boldmath $h$}^t\right)
   =
   \left(\mbox{\boldmath $\bar{K}$}\!\cdot\!\mbox{\boldmath
$Y$}\!\cdot\!
   \mbox{\boldmath $K$}^t\right),\label{Yij-lign}\\
   &&\left(\mbox{\boldmath $\bar K$}\!\cdot\!\mbox{\boldmath
$Y$}\!\cdot\!
   \mbox{\boldmath $K$}^t\right)\,
   \left(\mbox{\boldmath $\bar{ k}$}\!\cdot\!\mbox{\boldmath
$Y$}\!\cdot\!
   \mbox{\boldmath $k$}^t\right) -
   \left(\mbox{\boldmath $\bar{k}$}\!\cdot\!\mbox{\boldmath
$Y$}\!\cdot\!
   \mbox{\boldmath $K$}^t\right)\,
   \left(\mbox{\boldmath $\bar{K}$}\!\cdot\!\mbox{\boldmath
$Y$}\!\cdot\!
   \mbox{\boldmath $k$}^t\right)
   =
   \left(\mbox{\boldmath $\bar{H}$}\!\cdot\!\mbox{\boldmath
$X$}\!\cdot\!
   \mbox{\boldmath $H$}^t\right).
   \label{Xij-lign}
\eea
Regarding the expressions as polynomials in the indeterminates $H$,
$K$,
$\int K\,\partial H$  (with $\int H \partial K = H K -\int K\partial
H$),
and the corresponding barred quantities, (\ref{Xij-lign}) are
algebraic equations for {\boldmath $X$ and $Y$}.
Their solution is
\be
   \mbox{\boldmath $Y$} = \mbox{\boldmath $P$}\,
   \left(\mbox{\boldmath $X^t$}\right)^{-1}\,\mbox{\boldmath
$P$},\qquad
   \det\mbox{\boldmath $X$} = \det\mbox{\boldmath $Y$} = 1,
   \label{X12-restrictions}
\ee
where {\boldmath $P$} is the matrix
{\footnotesize
$\left(\begin{array}{rrr}0&0&1\\0&-1&0\\1&0&0\end{array}\right)$}.
We remind the reader that our definitions of $H,K$ and ${\bar
H},{\bar K}$
imply that
\be
  u = \left[(\partial H)^2\,\partial K\right]^{-1/3},\qquad
  v = \left[\partial H\,  (\partial K)^2\right]^{-1/3},
\ee
and the corresponding barred relations.
With these expressions, eq.\ (\ref{h12-form}), and the restrictions
(\ref{X12-restrictions}), we have found the general solution to
(\ref{eq:2.2}).
This solution is expressed in terms of the 4 arbitrary functions,
 $(H,K,{\bar H},{\bar K})$
and a $SL(3,R)$ matrix (say {\boldmath $X$, with $Y$ then been
determined as
the corresponding representation matrix for $X$}).

\subsection{The transformation group {\boldmath $SL(3,R)$}.}

The form of the solution we found above strongly indicates that there
is a
group structure behind it. As with the Liouville equation this can be
interpreted as the group of field transformations which leaves the
conservation laws invariant. It turns out that this group acts
linearely
on the fields $h$ and $k$.
First note that the conservation laws are related to the fields
$\varphi$ and $\chi$ through two $3rd$ and two $2nd$ order
differential
equations; thus there will appear a large number of integration
constants
in the solution of these equations. For a proper counting we find it
most
safe to (mentally) express everything in terms of the canonical
fields
$\varphi$, $\pi_{\varphi}$, $\chi$, $\pi_{\chi}$, and their spatial
derivatives. The number of integrations constants then turns out to
be
equal to the sum of the orders (spins) of the conservation laws minus
the
number of fields, here $2\times3+2\times2-2=8$.

Consider now our definitions (\ref{FG-vectors}) of {\boldmath $H$ and
$K$}.
With $u$ and $v$ given there is a freedom to shift $H$ and $K$ by
constants,
and the integral $\int K\,\partial H$ also contains
an undetermined integration constant (its lower limit of
integration). However,
such changes are equivalent to keeping {\boldmath $H$ and $K$ fixed,
and instead
multiplying $X$ and $Y$} by appropriate matrices. To be specific,
changing $H\to H+\alpha$ is equivalent to multiplying {\boldmath $X$
resp.\ $Y$}
from the right by
\be
  \mbox{e}^{\alpha F_1^{(X)}} = {\footnotesize
  \left(\begin{array}{ccc}
     1&0&0\\
\alpha&1&0\\
     0&0&1
  \end{array}\right)},\qquad
  \mbox{e}^{\alpha F_1^{(Y)}} = {\footnotesize
  \left(\begin{array}{ccc}
     1&0&0\\
     0&1&0\\
0&\alpha&1
  \end{array}\right)}.
\ee
Changing $K\to K + \beta$ is equivalent to multiplying {\boldmath $X$
resp.\ $Y$}
from the right by
\be
  \mbox{e}^{\beta F_2^{(X)}} = {\footnotesize
  \left(\begin{array}{ccc}
     1&0&0\\
     0&1&0\\
 0&\beta&1
  \end{array}\right)},\qquad
  \mbox{e}^{\beta F_2^{(Y)}} = {\footnotesize
  \left(\begin{array}{ccc}
     1&0&0\\
 \beta&1&0\\
     0&0&1
  \end{array}\right)}.
\ee
Changing $\int K\,\partial H \to \int K\,\partial H + \gamma$ is
equivalent to
multiplying {\boldmath $X$ resp.\ $Y$} by
\be
  \mbox{e}^{\gamma F_3^{(X)}} = {\footnotesize
  \left(\begin{array}{ccc}
     1&0&0\\
     0&1&0\\
\gamma&0&1
  \end{array}\right)},\qquad
  \mbox{e}^{\gamma F_3^{(Y)}} = {\footnotesize
  \left(\begin{array}{ccc}
      1&0&0\\
      0&1&0\\
-\gamma&0&1
  \end{array}\right)}.
\ee
There is a similar freedom to shift the corresponding barred
quantities. These shifts
are equivalent to keeping {\boldmath $\bar H$ and $\bar K$ unchanged,
and instead
multiplying $X$ resp.\ $Y$} from the left by the transpose of
matrices corresponding
to those above, i.e.\ exponentials of generators $E_i^{(\cdot)}$
which are
the transpose of the generators $F_i^{(\cdot)}$ above.

Our equations are also (for given $T$ and
$Q$) invariant under
scale transformations: $u\to \mbox{e}^{\mu}\,u$, $v\to
\mbox{e}^{\nu}\,v$,
$h\{u,v\}\to \mbox{e}^{\mu}\,h\{\mbox{e}^{\mu} u,\mbox{e}^{\nu} v\}$,
$k\{u,v\}\to \mbox{e}^{\nu}\,k\{\mbox{e}^{\mu} u,\mbox{e}^{\nu} v\}$.
Such transformations
are equivalent to keeping {\boldmath $H$ and $K$ fixed, and
multiplying $X$
resp.\ $Y$} from the right by
\be
   \mbox{e}^{\mu\,H_1^{(X)}+\nu\,H_2^{(X)}}={\footnotesize
    \left(\begin{array}{ccc}
    \mbox{e}^{\mu}&0&0\\
     0&\mbox{e}^{\nu-\mu}&0\\
     0&0&\mbox{e}^{-\nu}
    \end{array}\right)},
   \quad
   \mbox{e}^{\mu\,H_1^{(Y)}+\nu\,H_2^{(Y)}}={\footnotesize
    \left(\begin{array}{ccc}
    \mbox{e}^{\nu}&0&0\\
     0&\mbox{e}^{\mu-\nu}&0\\
     0&0&\mbox{e}^{-\nu}
    \end{array}\right)}
\ee
There is a similar invariance under rescalings of $\bar u$ and $\bar
v$. This
is equivalent to multiplications from the left by
the corresponding diagonal matrices.
The two sets of 8 generators
$\left\{ H_i^{(\cdot)}, E_j^{(\cdot)}, F_j^{(\cdot)} \right\}$
($i=1,2$, $j=1,\ldots,3$)
are representation matrices for a basis of the complex Lie algebra
$A_2$.
Upon exponentiation of these generators with real parameters, and
further composition,
they generate groups of real matrices, more specifically two copies
of $SL(3,R)$
which are (inequivalent) representations of each other.

The transformation on integration constants we have
considered above only corresponds to
the multiplication of {\boldmath $X$ and $Y$} by
lower triangular matrices from the right and upper triangular
matrices from the left. This does not generate the full group
structure, because there
are additional integration constants arising when we determine
$(u,v,\bar u,\bar v)$
from the conservation laws. But a change $u\to u'$ corresponds to a
multiplication of $h$ by the factor $u/u'$. Thus, as with the
Liouville
model, we expect that the freedom of changing all
integration constants are equivalent to multiplying {\boldmath $X$}
by
arbitrary elements of
$SL(3,R)$ from both the left and the right. A $SL(3,R)$ subgroup of
these
transformations are gauge transformations which leaves the
physical fields unchanged.

\subsection{The initial value problem.}

Let us summarize how the initial value problem can be solved
using the results above:

\begin{enumerate}

\item From the initial data at time $t=t_0$
one computes the conserved densities
$T,{\overline T}, Q, {\overline Q}$. Note that this process
requires a rewriting of the quantities in terms of
the canonical fields and their spatial derivatives.
This can be done with the use of
the equations of motion, although the expressions become rather
lengthy.

\item
{}From the knowledge of the conserved densities one determines
$\left\{u,v,{\bar u},{\bar v}\right\}$, using (\ref{V12-lign}) with
the appropriate
replacements mentioned above. For most initial data these equations
will have to be solved numerically. Note that we may choose a fixed
set of initial conditions in this process.

\item
With further integrations one finds the vectors
$\left\{
\mbox{\boldmath $H$},
\mbox{\boldmath $K$},
\mbox{\boldmath ${\bar H}$},
\mbox{\boldmath ${\bar K}$}
\right\}$.
We may also choose a fixed set of integration constants
in this process.

\item
Finally {\boldmath $X$ and the corresponding $Y$} aare determined so
that
the solutions $\varphi = -\log(\bar u\,h\,u)$ and $\chi=-\log(\bar
v\,k\,v)$
reproduce the initial values of
the canonical fields (and a sufficient number of their spatial
derivatives)
at some initial point $(t_0,x_0)$.

\end{enumerate}

\section{Solution of {\boldmath $B_2$} Toda field theory}

The $B_2$ Toda field theory can be defined by the Lagrangian
\be
    {\cal L} = \barpartial\phi_1\,\partial\phi_1 +
               \barpartial\phi_2\,\partial\phi_2
   -\frac{1}{2}\sigma_1\mbox{e}^{2(\phi_1-\phi_2)} -
\sigma_2\,\mbox{e}^{2\phi_2},
   \label{B_2-lag}
\ee
where the $\sigma_i$'s are parameters which may be set to $\pm1$
after
appropriate redefinitions. Only the case
$\sigma_1=\sigma_2=1$ gives a Hamiltonian which is bounded from
below. We
shall assume this case in our further analysis. The equations of
motion simplify
if we introduce new fields $\varphi=2\phi_1$ and
$\chi=\phi_1+\phi_2$, in which case
they become
\be
\partial\barpartial\varphi
=
-\sigma\, \mbox{e}^{2\varphi-2\chi},
\qquad
\partial\barpartial\chi
=
-\sigma\, \mbox{e}^{-\varphi+2\chi}.
\label{eq:3.2}
\ee
These equations have no apparent symmetry. However, the ansatz
\be
  \varphi = 2\,\phi + \log 3,\qquad
  \chi = \frac{3}{2}\,\phi + \log (3/\sqrt{2}),
  \label{redux}
\ee
is consistent with the
equations, and leads to the Liouville equation for $\phi$.

\subsection{General solution}

Equations (\ref{eq:3.2}) also have two sets of basic conserved
densities.
With the notation that
$\varphi_n\equiv\partial^n\varphi$, $\chi_n\equiv\partial^n\chi$ the
set corresponding to the conformally improved  energy momentum tensor
is
\be
T =
\frac{1}{2}\,\varphi_1^2
-\varphi_1\chi_1
+\chi_1^2
-\frac{1}{2}\,\varphi_2
-\chi_2,
\label{eq:3.3}
\ee
and the same expression for $\overline T$ with the replacement
$\partial\to\barpartial$.
Eq.\ (\ref{eq:3.2}) admit no spin 3 conserved densities, but there is
a set of spin 4
conserved densities\cite{HohlerOlaussen2},
\bea
Q
&=&
\varphi_1^4
-4\,\varphi_1^3\,\chi_1
+12\,\varphi_1^2\,\chi_1^2
-16\,\varphi_1\,\chi_1^3
+8\,\chi_1^4
-2\,\varphi_1^2\,\varphi_2
-4\,\varphi_1\,\chi_1\,\varphi_2
\nonumber\\*
&&
+3\,\varphi_2^2
-8\,\varphi_1^2\,\chi_2
+24\,\varphi_1\,\chi_1\,\chi_2
-16\,\chi_1^2\,\chi_2
+4\,\varphi_2\,\chi_2
+4\,\chi_2^2
\label{eq:3.4}\\*
&&
+2\,\varphi_1\,\varphi_3
+2\,\chi_1\,\varphi_3
-2\,\varphi_1\,\chi_3
-4\,\chi_1\,\chi_3
-\varphi_4
+2\,\chi_4,
\nonumber
\eea
and the same expression for $\overline Q$ with the replacement
$\partial\to\barpartial$. These conserved densities satisfy
$\barpartial T = \partial{\overline T} =
\barpartial Q=\partial{\overline Q}=0$.
If we insert the ansatz (\ref{redux}) into these
expressions the result must be expressible in terms of the
conservation laws
of the Liouville equation. Indeed we find
\(
   T = \frac{5}{2}\,T_{L}
\),
and \(
   Q = \frac{136}{25}\,T_{L}^2 -
   \frac{2}{5}\,\partial^2 T_{L}
\),
where $T_{L}$ is the energy-momentum tensor (\ref{ImprovedTensor}) of
the Liouville model.
This provides a check of our expressions.

It is convenient to introduce new variables $U =\mbox{e}^{-\varphi}$
and $V=\mbox{e}^{-\chi}$. We obtain, with the notation that
$U_n\equiv\partial^n U$ and
$V_n\equiv \partial^n V$,
\be
  T = \frac{1}{2}\,\frac{U_2}{U} + \frac{V_2}{V} -
\frac{U_1\,V_1}{U\,V},
  \label{B2-T}
\ee
and a lengthy expression for $Q$. It is possible to use (\ref{B2-T})
to eliminate
all explicit reference to $V$ resp.\ $U$ in the expression for $Q$.
We get
\be
  \frac{U_4}{U} - \frac{U_3\,U_1}{U^2} +
  \frac{1}{2}\,\frac{U_2^2 }{U^2} -4\,T\,\frac{U_2}{U} +
  2\,T\,\frac{U_1^2}{U^2} - 2\,\partial T\,\frac{U_1}{U}
  = \frac{1}{2}\,Q + \partial^2 T - 4\, T^2
  \label{V1-lign}
\ee
and
\be
   \frac{V_4}{V} -2\,T\,\frac{V_2}{V}
   -2\,\partial T\,\frac{V_1}{V} =
   T^2 -\frac{1}{2}\, \partial^2 T - \frac{1}{4}\,Q.
   \label{V2-lign}
\ee
The latter is seen to be a linear $4th$ order equation for $V$.

As before we write $U = {\bar u}(\zbar)\,h(z,\zbar)\,u(z)$,
$V = {\bar v}(\zbar)\,k(z,\zbar)\,v(z)$,
and make the ansatz that $T$ and $Q$ are functionals of $u$ and $v$
only (and that
$\overline T$ and $\overline Q$ are functionals of $\bar u$ and $\bar
v$ only).
This means that $u$ must satisfy (\ref{V1-lign}) with the replacement
$U \to u$, and
$v$ must satisfy (\ref{V2-lign}) with the replacement $V \to v$. With
these assumptions we find equations for $h$ and $k$. The equation for
$h$ can be written
in the form
\be
  \left(\frac{u^2}{v^2}\,\partial\,\frac{v^2}{u} \,\partial\,
  \frac{v^2}{u} \,\partial\, \frac{u^2}{v^2} \,\partial h\right)\,h +
R = 0,
  \label{h-nonlin}
\ee
where $R$ is rather complicated. With the notation that
$h_n\equiv\partial^n h$,
$u_n\equiv\partial^n u$, and $v_n\equiv\partial^n v$:
\be
   R = -u^2 h_1 h_3 + \frac{1}{2} u^2 h_2^2
   - u u_1 h_1 h_2 +
   \left(2 u_1^2-2 u u_2 -
    \frac{u^2}{v} v_2 - \frac{u}{v} u_1 v_1
   \right) h_1^2.
\ee
However, the derivative of $R$ has a compact representation,
\be
   \partial R = -\left(\frac{u^2}{v^2} \,\partial\, \frac{v^2}{u}
\,\partial\,
   \frac{v^2}{u} \,\partial\, \frac{u^2}{v^2} \,\partial\, h\right)
\partial h.
\ee
Thus, by differentiating (\ref{h-nonlin}) we find that $h$ satisfies
a
$5th$ order linear equation,
\be
   \partial\,\frac{ u^2}{v^2} \,\partial\, \frac{v^2}{u}\,\partial\,
   \frac{v^2}{u}\,\partial\,\frac{u^2}{v^2} \,\partial\, h = 0.
   \label{h-lign}
\ee
The equation for $k$ can be written as
\be
    \partial\,\frac{v^2}{u} \,\partial\, \frac{u^2}{v^2}
    \,\partial \,\frac{v^2}{u} \,\partial k = 0.
   \label{k-lign}
\ee
It is now straigthforward to integrate (\ref{h-lign}) and
(\ref{k-lign}).
Define functions $H_1$ and $K_1$ such that $\partial H_1 = v^2/u^2$
and
$\partial K_1 = u/v^2$.
Then
\bea
   h &=& \bar\alpha_0 + \bar\alpha_1\,H_1 + \bar\alpha_2\,H_2
         + \bar\alpha_3\,H_3 + \bar\alpha_4\, H_4,\nonumber\\
   k &=& \bar\beta_0 + \bar\beta_1\,K_1 + \bar\beta_2\,K_2 +
\bar\beta_3\,K_3,
   \label{h_and_k}
\eea
where all the coefficients $\bar\alpha_i$, $\bar\beta_i$
may depend on $\zbar$: $\bar\alpha_0 = \bar\alpha_0(\zbar)$ etc. The
functions
$H_i$ and $K_i$ are given by the integrals
\bea
  H_2 &=& \int \partial H_1 \int \partial K_1\nonumber =
          \int K_1\,\partial H_1,\\
  H_3 &=& \int \partial H_1 \int \partial K_1 \int \partial K_1 =
          \frac{1}{2}\,\int K_1^2\, \partial H_1\nonumber,\\
  H_4 &=& \int \partial H_1 \int \partial K_1 \int \partial K_1 \int
\partial H_1 =
          H_1\,H_3 - \frac{1}{2}\,H_2^2,\label{Fs_and_Gs}\\
  K_2 &=& \int \partial K_1 \int \partial H_1 =
          \int H_1\,\partial K_1 = H_1\,K_1 - H_2,\nonumber\\
  K_3 &=& \int \partial K_1 \int \partial H_1 \int \partial K_1
          = H_2\,K_1 - 2\,H_3.\nonumber
\eea
Here the various identities among the integrals are derived by
(repeated) partial
integrations, and depends on certain assumptions about the
undetermined integration
constants (i.e., the lower integration limits). This is the maximum
set
of relations which may be derived between the integrals
without making specific assumptions about
the functions $u$ and $v$. It is also the minumum set required for
the rest of
the solution to go through.
However, the identities still allow
us to freely change the outer integration limits in the definitions
of $H_2$ and $H_3$,
and to shift $H_1$ and $K_1$ by independent constants.
Note that we have the relation
$2\,H_4-2\,H_3 H_1 + H_2^2 =0$. With the additional constraint that
$2\,{\bar\alpha}_4{\bar\alpha}_0
-2\,{\bar\alpha}_3{\bar\alpha}_1+{\bar\alpha}_2^2=0$ the
expression (\ref{h_and_k}) for $h$ is also a solution of
(\ref{h-nonlin}),
which in fact was the equation we first solved to determine $h$.

The functions $h$ and $k$ must also satisfy differential equations in
$\zbar$, the same
as (\ref{h-lign}) and (\ref{k-lign}) with the replacements
$(u,v,\partial)\to(\bar u,\bar v,\barpartial)$. This means that we
also have the
representation
\bea
   h &=& \alpha_0 + \alpha_1\,{\bar H}_1 + \alpha_2\,{\bar H}_2
         + \alpha_3\,{\bar H}_3 + \alpha_4\, {\bar H}_4,\nonumber\\
   k &=& \beta_0 + \beta_1\,{\bar K}_1 + \beta_2\,{\bar K}_2 +
\beta_3\,{\bar K}_3,
   \label{h_and_k1}
\eea
where the $\bar H_i$'s and $\bar K_i$'s are given by the barred
version of
(\ref{Fs_and_Gs}),
and the coefficients $\alpha_i$, $\beta_i$ may depend on $z$:
$\alpha_0= \alpha_0(z)$ etc. Equations (\ref{h_and_k},\ref{h_and_k1})
are consistent
with each other if each of the coefficients $\alpha_i$ is a linear
combination of
the functions $1$, $H_1$, $H_2$, $H_3$, $H_4$, and each of the
coefficients
$\beta_i$ is a linear combination of the functions $1$, $K_1$, $K_2$,
$K_3$. Define
(with coefficients choosen for later convenience)
\bea
   &\mbox{\boldmath $H$} = (1,H_1,2^{1/2}\,H_2,2\,H_3,2\,H_4),\qquad
   &\mbox{\boldmath $K$} = (1,K_1,K_2,K_3)\\
   &\mbox{\boldmath $\bar H$} = (1,\bar H_1,2^{1/2}\,\bar H_2,2\,
         \bar H_3,2\,\bar H_4),\qquad
   &\mbox{\boldmath $\bar K$} = (1,\bar K_1,\bar K_2,\bar K_3).
\eea
Then all the linear equations are satisfied by the representations
\be
   h = \mbox{\boldmath $\bar H$}\!\cdot\!\mbox{\boldmath
$X$}\!\cdot\!
       \mbox{\boldmath $H$}^t,
   \qquad
   k = \mbox{\boldmath $\bar K$}\!\cdot\!\mbox{\boldmath
$Y$}\!\cdot\!
   \mbox{\boldmath $K$}^t,
   \label{h_and_k2}
\ee
where {\boldmath $X$} is a $5\times5$ matrix of real constants and
{\boldmath $Y$} is
a $4\times4$ matrix of real constants. However, not all the
representations
(\ref{h_and_k2}) will satisfy the original Toda equations
(\ref{eq:3.2}). To check our
ansatz we first compute
\bea
   &&\partial \mbox{\boldmath $H$} =
   (0,1,2^{1/2}\,K_1, K_1^2, H_1\,K_1^2 - 2\,H_2\,K_1 + 2\,H_3)
   \,\partial H_1 \equiv \mbox{\boldmath $h$}\,\partial H_1,
   \nonumber\\&&
   \partial \mbox{\boldmath $K$} =
   (0,1,H_1,H_2)\,\partial K_1 \equiv \mbox{\boldmath $h$}\,\partial
K_1,
\eea
and the corresponding barred quantities. Inserting
$\varphi=-\log({\bar u}\,h\,u)$ and $\chi=-\log({\bar v}\,k\,v)$
into (\ref{eq:3.2}) gives
the equations $h\,\barpartial\partial h  -\barpartial h\,\partial h =
k^2\,\barpartial{\bar H}_1\,\partial H_1$ and
$k\,\barpartial\partial k  -\barpartial k\,\partial k =
h\,\barpartial{\bar K}_1\,\partial K_1$. These are purely algebraic
conditions\footnote{If one wants to solve the model (\ref{B_2-lag})
with
$\sigma_1,\sigma_2\in \left\{1,-1\right\}$ one should at this point
multiply the right hand side of (\ref{phi-lign}) by $\sigma_1$ and
the
right hand side of (\ref{chi-lign}) by $\sigma_2$.
}:
\bea
   &&\left(\mbox{\boldmath $\bar H$}\!\cdot\!\mbox{\boldmath
$X$}\!\cdot\!
   \mbox{\boldmath $H$}^t\right)\,
   \left(\mbox{\boldmath $\bar h$}\!\cdot\!\mbox{\boldmath
$X$}\!\cdot\!
   \mbox{\boldmath $h$}^t\right) -
   \left(\mbox{\boldmath $\bar h$}\!\cdot\!\mbox{\boldmath
$X$}\!\cdot\!
   \mbox{\boldmath $H$}^t\right)\,
   \left(\mbox{\boldmath $\bar H$}\!\cdot\!\mbox{\boldmath
$X$}\!\cdot\!
   \mbox{\boldmath $h$}^t\right)
   =
   \left(\mbox{\boldmath $\bar K$}\!\cdot\!\mbox{\boldmath
$Y$}\!\cdot\!
   \mbox{\boldmath $K$}^t\right)^2,
   \label{phi-lign}\\
   &&\left(\mbox{\boldmath $\bar K$}\!\cdot\!\mbox{\boldmath
$Y$}\!\cdot\!
   \mbox{\boldmath $K$}^t\right)\,
   \left(\mbox{\boldmath $\bar k$}\!\cdot\!\mbox{\boldmath
$Y$}\!\cdot\!
   \mbox{\boldmath $k$}^t\right) -
   \left(\mbox{\boldmath $\bar k$}\!\cdot\!\mbox{\boldmath
$Y$}\!\cdot\!
   \mbox{\boldmath $K$}^t\right)\,
   \left(\mbox{\boldmath $\bar {K}$}\!\cdot\!\mbox{\boldmath
$Y$}\!\cdot\!
   \mbox{\boldmath $k$}^t\right)
   =
   \left(\mbox{\boldmath $\bar {H}$}\!\cdot\!\mbox{\boldmath
$X$}\!\cdot\!
   \mbox{\boldmath $H$}^t\right).
   \label{chi-lign}
\eea
Regarding the expressions as polynomials in the indeterminates
$H_1$, $H_2$, $H_3$, $K_1$ (and the corresponding barred quantities)
these are sets of algebraic equations for
the elements of {\boldmath $X$ and $Y$}.

\subsection{The transformation group {\boldmath $Sp(2,R)$}}

The equations (\ref{phi-lign},\ref{chi-lign}) are nonlinear and
involve
many ($5^2+4^2=41$) variables. They are not easy to solve by brute
force.
However, with the ansatz that {\boldmath $X$ and $Y$} are diagonal a
two-parameter class of solutions is straightforward to find:
\be
   \mbox{\boldmath $X$} = \mbox{diag}\left(\mbox{e}^{\mu},
\mbox{e}^{2\nu-\mu},1,
\mbox{e}^{\mu-2\nu},\mbox{e}^{-\mu}\right),\qquad
   \mbox{\boldmath $Y$} = \mbox{diag}\left(\mbox{e}^{\nu},
    \mbox{e}^{\mu-\nu},\mbox{e}^{\nu-\mu},\mbox{e}^{-\nu}
    \right).
   \label{diag-sol}
\ee
The two parameters have their origin in the symmetry
that our equations for given $T$ and
$Q$ are invariant under: $u\to \mbox{e}^{\mu}\,u$, $v\to
\mbox{e}^{\nu}\,v$,
$h\{u,v\}\to \mbox{e}^{\mu}\,h\{\mbox{e}^{\mu} u,\mbox{e}^{\nu} v\}$,
$k\{u,v\}\to \mbox{e}^{\nu}\,k\{\mbox{e}^{\mu} u,\mbox{e}^{\nu} v\}$.
Such transformations
are equivalent to keeping {\boldmath $H$ and $K$ fixed, and
multiplying $X$
resp.\ $Y$} from the right by diagonal matrices,
\bea
   \exp\left(\mu\,H_1^{(X)}+\nu\,H_2^{(X)}\right)&=&\mbox{diag}
\left(\mbox{e}^{\mu},\mbox{e}^{2\nu-\mu},1,
\mbox{e}^{\mu-2\nu},\mbox{e}^{-\mu}\right),
   \quad\nonumber\\
   \exp\left(\mu\,H_1^{(Y)}+\nu\,H_2^{(Y)}\right)&=&\mbox{diag}
\left(\mbox{e}^{\nu},\mbox{e}^{\mu-\nu},
\mbox{e}^{\nu-\mu},\mbox{e}^{-\nu} \right).
\eea
There is a similar invariance under rescalings of $\bar u$ and $\bar
v$. This
is equivalent to multiplications from the left by the corresponding
diagonal matrices.

{}From the solutions (\ref{diag-sol}) we can easily find many more,
since the lower integration limits in the expressions
(\ref{Fs_and_Gs}) for $H_1$, $H_2$, $H_3$ and $K_1$ were not fully
specified.
We are free to shift $H_1\to H_1+\alpha$. This is equivalent to
keeping
{\boldmath $H$ and $K$ unchanged, and instead multiplying
$X$ resp.\ $Y$} from the right by matrices
\[
  \mbox{e}^{\alpha\,F_1^{(X)}} =
  \footnotesize{\left(\begin{array}{ccccc}
               1&0&0&0&0\\
          \alpha&1&0&0&0\\
               0&0&1&0&0\\
               0&0&0&1&0\\
               0&0&0&\alpha&1
 \end{array}\right)},\quad
 \mbox{e}^{\alpha\,F_1^{(Y)}} =
 \footnotesize{\left(\begin{array}{cccc}
               1&0&0&0\\
               0&1&0&0\\
               0&\alpha&1&0\\
               0&0&0&1
 \end{array}\right)}
\]
We are free to shift $K_1\to K_1+\beta$. This is equivalent to
keeping
{\boldmath $H$ and $K$ unchanged, and instead multiplying
$X$ resp.\ $Y$} from the right by matrices
\[
  \mbox{e}^{\beta\,F_2^{(X)}} =
\footnotesize{\left(\begin{array}{ccccc}
               1&0&0&0&0\\
               0&1&0&0&0\\
         0&\beta\sqrt{2}&1&0&0\\
         0&\beta^2&\beta\sqrt{2}&1&0\\
               0&0&0&0&1
 \end{array}\right)},\quad
 \mbox{e}^{\beta\,F_2^{(Y)}} =
\footnotesize{\left(\begin{array}{cccc}
                 1&0&0&0\\
             \beta&1&0&0\\
                 0&0&1&0\\
             0&0&\beta&1
 \end{array}\right)}
\]
We are free to shift $H_2\to H_2+\gamma$. This is equivalent to
keeping
{\boldmath $H$ and $K$ unchanged, and instead multiplying
$X$ resp.\ $Y$} from the right by matrices
\[
  \mbox{e}^{\gamma\,F_3^{(X)}} =
  \footnotesize{\left(\begin{array}{ccccc}
               1&0&0&0&0\\
               0&1&0&0&0\\
          \gamma\sqrt{2}&0&1&0&0\\
               0&0&0&1&0\\
        -\gamma^2&0&-\gamma\sqrt{2}&0&1
 \end{array}\right)},\quad
 \mbox{e}^{\gamma\,F_3^{(Y)}} =
\footnotesize{\left(\begin{array}{cccc}
               1&0&0&0\\
               0&1&0&0\\
             -\gamma&0&1&0\\
             0&{\gamma}&0&1
 \end{array}\right)}
\]
We are free to shift $H_3\to H_3 - \delta/2$. This is equivalent to
keeping
{\boldmath $H$ and $K$ unchanged, and instead multiplying
$X$ resp.\ $Y$} from the right by matrices
\[
  \mbox{e}^{\delta\,F_4^{(X)}} =
  \footnotesize{\left(\begin{array}{ccccc}
               1&0&0&0&0\\
               0&1&0&0&0\\
               0&0&1&0&0\\
          -\delta&0&0&1&0\\
          0&-\delta&0&0&1
 \end{array}\right)},\quad
  \mbox{e}^{\delta\,F_4^{(Y)}} =
\footnotesize{\left(\begin{array}{cccc}
               1&0&0&0\\
               0&1&0&0\\
               0&0&1&0\\
         \delta&0&0&1
 \end{array}\right)}
\]
There is a similar freedom to shift the corresponding barred
quantities. These shifts
are equivalent to keeping {\boldmath $\bar H$ and $\bar K$ unchanged,
and instead
multiplying $X$ resp.\ $Y$} from the left by the transpose of
matrices corresponding
to those above, i.e.\ exponentials of generators $E_i^{(\cdot)}$
which are the transpose
of the generators $F_i^{(\cdot)}$ above. The two sets of 10
generators
$\left\{ H_i^{(\cdot)}, E_j^{(\cdot)}, F_j^{(\cdot)} \right\}$
($i=1,2$, $j=1,\ldots,4$)
are representation matrices for a basis of the complex Lie algebra
$B_2$.
Upon exponentiation of these generators with real parameters, and
further composition,
they generate groups of real matrices. More specifically, with a
symplectic form
{\boldmath $\epsilon$ and a metric form $\eta$},
\be
    \mbox{\boldmath $\epsilon$}=
{\footnotesize\left(\begin{array}{cccc}
     0&0& 0&1\\
     0&0&-1&0\\
     0&1& 0&0\\
    -1&0& 0&0
   \end{array}\right)},\qquad
    \mbox{\boldmath $\eta$}= {\footnotesize\left(\begin{array}{ccccc}
    0& 0&0& 0&1\\
    0& 0&0&-1&0\\
    0& 0&1& 0&0\\
    0&-1&0& 0&0\\
    1& 0&0& 0&0
   \end{array}\right)},
\ee
the 4-dimensional representation generate real symplectic matrices
$\mbox{\boldmath $S$} \in Sp(2,R)$
({\boldmath $S^t\,\epsilon\,S = \epsilon$}),
and the 5-dimensional representation generate pseudo-orthogonal
matrices
$\mbox{\boldmath $O$} \in O(3,2)$
({\boldmath $O^t\,\eta\,O = \eta$}).
It is now straightforward to verify by direct insertion that a
solution to
(\ref{phi-lign},\ref{chi-lign}) is obtained by choosing any
$\mbox{\boldmath $Y$}\in Sp(2,R)$ (as defined above) and taking
{\boldmath $X$} to
be the corresponding $SO(3,2)$ representation matrix. The precise
connection between
corresponding {\boldmath $X$ and $Y$}'s is implicitly defined by the
connection
between the 4- and 5-dimensional generators listed in the appendix.
A more direct relation between {\boldmath $X$ and $Y$} is defined by
(\ref{chi-lign}), which moreover is an explicit expression for the
function $h$.
Eq.\ (\ref{chi-lign}) show that {\boldmath $Y$} and $-\mbox{\boldmath
$Y$}$ leads
to the same solution for {\boldmath $X$}, which means
that $Sp(2,R)$ is a double covering of $SO(3,2)$. Thus the
transformation group
in this case is $Sp(2,R)$ (or rather
$Sp(2,R)\times Sp(2,R)$ if we include the gauge transformations).

The initial value problem can now be solved by the same procedure as
described for
the $A_2$ Toda field theory. The practical implementation of this
procedure only
becomes more cumbersome.

\section{Concluding remarks}

In this paper we have in detail discussed the explicit solutions of
the three
simplest Toda field theories. From this a general picture emerges,
which
seems to be as follows:
For each Lie algebra $A_n$ ($n\ge1$), $B_n$ ($n\ge2$), $C_n$
($n\ge3$),
$D_n$ ($n\ge3$), $E_n$ ($n=6,7,8$), $F_4$, and $G_2$
there exists a Toda field theory.
The number of independent field components in the model
is equal to the index $n$. Thus, to solve the model by the method
used in
this paper we have to find $2n$ independent conservation laws which
can be written
in one-sided forms,
\be
       \barpartial Q_s = 0,\qquad \partial{\overline Q}_s = 0.
\ee
Such conservation laws can be found in all the cases we have
considered.
The conserved densities
always occur in parity related
pairs $(Q_s,{\overline Q}_s)$.
For $A_n$ there is one independent density $Q_s$
for each spin $s=2,3,\ldots,n+1$. For $B_n$ and $C_n$ there is one
independent density $Q_s$
for each spin $s=2,4,\ldots,2n$.
For $D_n$ there is one independent density $Q_s$
for each spin $s=2,4,\ldots,2n-2$, and an additional one of spin
$s=n$.
For $E_6$ there is one independent density for each spin
$s\in\left\{ 2,5,6,8,9,12 \right\}$.
For $E_7$ there is one independent
density for each spin $s\in\left\{ 2,6,8,10,12,14,18 \right\}$.
For $E_8$ there is one independent
density for each spin $s\in\left\{ 2,8,12,14,18,20,24,30 \right\}$.
For $F_4$ there is one independent
density for each spin $s\in\left\{ 2,6,8,12 \right\}$.
For $G_2$ there is one independent
density for each spin $s\in\left\{ 2,6\right\}$. Counting shows that
the number of independent conserved densities is always equal to the
number of
field components.

The fields in the model are related to a conserved density
by a differential relation of the same order as its spin.
By considering the canonical degrees of freedom one finds that
the number $N$ of integration constants which occur when one attempts
to determine the canonical fields from the conserved densities is
equal to the sum of the spins (counting both $Q_s$ and
${\overline Q}_s$), minus the number of field components.
This number is $N=(n+1)^2-1$ for $A_n$,
$N=(2n+1)n$ for $B_n$ and $C_n$, $N=(2n-1)n$ for $D_n$,
and $N=(78,133,248,52,14)$ for $(E_6,E_7,E_8,F_4,G_2)$.
These numbers are precisely equal to the dimensions of the Lie
algebras which label the models. Thus there is a $N$-dimensional
manifold of
field configurations which lead to the same conserved densities, and
a
continuous transformation group acting on this manifold. Our
experience
from the previous sections is that the manifold may
be identified with the transformation group, and that the group has
as
its Lie algebra one of the real forms of the (complex) Lie algebra
which labels the Toda field theory.
The freedom of choosing integration constants is probably
such that the transformation group will consist of one simply
connected component.

How useful are the solutions we have found? If the purpose is to
determine
the classical fields on a dense grid of points covering a
finite regular region of space-time, it may
in fact be difficult to beat a direct numerical solution of the
partial
differential equations in efficiency (although probably in accuracy).
The computational effort must in any case be proportional to the
number of
grid points, and the proportionality factor is rather small for a
direct
numerical method.
However, if we are interested in finding the fields along a line or
on small set of points only, the use of our solutions would be vastly
more
efficient than a direct numerical method. The same is true if the
interest
is in how asymptotic field configurations at $t\to-\infty$
develop into asymptotic field configurations at $t\to\infty$.
Finally there is the problem of understanding the quantized versions
of these models. For this any improved understanding of their
analytic structure
is of potential use.

\appendix

\section{The complex Lie algebra {\boldmath $B_2$}.}

The complex Lie algebra $B_2$ can be represented by $5\times5$
or $4\times4$ matrices. Let
$\mbox{diag}^{n\uparrow}\left(\cdot,\ldots,\cdot\right)$
denote a matrix whose only non-zero entries are on the $n$'th
superdiagonal,
and $\mbox{diag}^{n\downarrow}\left(\cdot,\ldots,\cdot\right)$ denote
a matrix whose
only non-zero entries are on the $n$'th subdiagonal. Then the
transformations
found in section 4.2  correspond to a 5-dimensional
matrix representation of a basis of $B_2$,
\be
\begin{array}{lcl}
   H_1^{(5)}=\mbox{diag}\left(1,-1,0,1,-1\right),&\quad
   &H_2^{(5)}=\mbox{diag}\left(0,2,0,-2,0\right),\\
   E_1^{(5)}=\mbox{diag}^{1\uparrow}\left(1,0,0,1\right),&\quad

&E_2^{(5)}=\mbox{diag}^{1\uparrow}\left(
0,\sqrt{2},\sqrt{2},0\right),\\

E_3^{(5)}=\mbox{diag}^{2\uparrow}\left(
\sqrt{2},0,-\sqrt{2}\right),&\quad
   &E_4^{(5)}=\mbox{diag}^{3\uparrow}\left(-1,-1\right),\\
   F_1^{(5)}=\mbox{diag}^{1\downarrow}\left(1,0,0,1\right),&\quad

&F_2^{(5)}=\mbox{diag}^{1\downarrow}\left(0,
\sqrt{2},\sqrt{2},0\right),\\

F_3^{(5)}=\mbox{diag}^{2\downarrow}
\left(\sqrt{2},0,-\sqrt{2}\right),&\quad
   &F_4^{(5)}=\mbox{diag}^{3\downarrow}\left(-1,-1\right),
\end{array}
\ee
and a corresponding 4-dimensional representation,
\bea
\begin{array}{lcl}
   H_1^{(4)}=\mbox{diag}\left(0,1,-1,0\right),&\qquad\quad
   &H_2^{(4)}=\mbox{diag}\left(1,-1,1,-1\right),\\
   E_1^{(4)}=\mbox{diag}^{1\uparrow}\left(0,1,0\right),&\qquad\quad
   &E_2^{(4)}=\mbox{diag}^{1\uparrow}\left(1,0,1\right),\\
   E_3^{(4)}=\mbox{diag}^{2\uparrow}\left(-1,1\right),&\qquad\quad
   &E_4^{(4)}=\mbox{diag}^{3\uparrow}\left(1\right),\\
   F_1^{(4)}=\mbox{diag}^{1\downarrow}\left(0,1,0\right),&\qquad\quad
   &F_2^{(4)}=\mbox{diag}^{1\downarrow}\left(1,0,1\right),\\
   F_3^{(4)}=\mbox{diag}^{2\downarrow}\left(-1,1\right),&\qquad\quad
   &F_4^{(4)}=\mbox{diag}^{3\downarrow}\left(1\right).
\end{array}
\eea
These generators are given in the Chevalley basis. Thus they are
normalized so that
$\left[E_i,F_j\right] = \delta_{ij}\,H_i$ for
$i,j\in\left\{1,2\right\}$.
Not all relative signs are determined by the conventions of a
Chevalley basis. Here we have chosen signs so that
$[E_1,E_2]=  E_3$ and $[E_2,E_3]=2 E_4$.

%\include{a2solur}
% commands for the bibliography
\newcommand{\Author}[1]{#1,}
\newcommand{\Title}[1]{{\em #1},}
\newcommand{\Journal}[1]{#1}


\begin{thebibliography}{99}

\bibitem{Arnold}
\Author{J.\ Liouville}
\Journal{Jour.\ de Math.\ {\bf 20} (1855) 137.
See e.g.~V.~I.~Arnold,
{\em Mathematical Methods of Classical Mechanics, Ch.~10}, 271--278,
Springer-Verlag 1979.}

\bibitem{Bogoyavlenski}
\Author{O.I.\ Bogoyavlenski}
\Title{On perturbations of the periodic Toda lattice}
\Journal{Comm.\ Math.\ Phys.\ {\bf 51} (1976) 201--209}

\bibitem{LeznovSaveliev}
\Author{A.N.\ Leznov and M.V.\ Saveliev}
\Title{Representation of zero curvature for the system of
nonlinear partial differential equations $x_{\alpha,z\zbar} =
\exp\left(kx\right)_{\alpha}$ and its integrability}
\Journal{Lett.\ Math.\ Phys.\ {\bf 3} (1979) 489--494}

\bibitem{BilalGervais1}
\Author{A.\ Bilal and J.-L.\ Gervais}
\Title{Systematic approach to conformal systems with extended
Virasoro
symmetries}
\Journal{Phys.\ Lett.\ B {\bf 206} (1988) 412--420}

\bibitem{BilalGervais2}
\Author{A.\ Bilal and J.-L.\ Gervais}
\Title{Extended $C=\infty$ conformal systems from classical Toda
field
theories}
\Journal{Nucl.\ Phys.\ B\ {\bf 314} (1989) 646--686}

\bibitem{Mansfield}
\Author{P.\ Mansfield}
\Title{Solution of Toda systems}
\Journal{Nucl.\ Phys. B\ {\bf 257}, (1982) 277--300}

\bibitem{Palla}
\Author{Z.\ Bajnok, L.\ Palla and G.\ Tak\'acs}
\Title{$A_2$ Toda theory in the reduced WZNW framework and the
representations of the $W$ algebra}
\Journal{Nucl.\ Phys. B\ {\bf 385}, (1992) 329--360}

\bibitem{Underwood}
\Author{J.\ Underwood}
\Title{Aspects of non-Abelian Toda theories}
\Journal{hep-th/9304156}

\bibitem{Liouville}
\Author{J.\ Liouville}
\Title{Sur l'equation aux diff\'erences partielles}
\Journal{J.\ Math.\ Pures et Appliques 18, (1853) 71--72}.
See e.g.\
\Author{P.G.Drazin and R.S.Johnson}
\Title{Solitons: An Introduction}
\Journal{Cambridge Texts in Applied Mathematics, Cambridge University
Press (1989)},
or
\Author{G.L.Lamb Jr.}
\Title{Elements of Soliton Theory}
\Journal{John Wiley \& Sons (1980)}

\bibitem{PapaSpence}
\Author{G.\ Papadopoulos and B.\ Spence}
\Title{A covariant canonical description of Liouville field theory}
\Journal{Phys.\ Lett.\ B {\bf 308} (1993) 253--259}

\bibitem{HohlerOlaussen1}
\Author{E.G.B.\ Hohler and K.\ Olaussen}
\Title{Conservation laws for the classical Toda field theories}
\Journal{Mod.\ Phys.\ Lett.\ A 8 (1993) 3377-3385}
and \Journal{hep-th/9404069}

\bibitem{HohlerOlaussen2}
\Author{E.G.B.\ Hohler and K.\ Olaussen}
\Title{Explicit calculation of conservation laws for Toda field
theories}
\Journal{Preprint, Theoretical Physics Seminar in Trondheim, No 8,
1995}

\end{thebibliography}
\end{document}